\documentclass[epjST]{svjour}

\usepackage{graphicx}
\usepackage{hyperref}
\usepackage{amsmath}
\usepackage{color}

\begin{document}

\title{Modeling the Multi-layer Nature of the European Air Transport Network: Resilience and Passengers Re-scheduling under random failures}
\author{Alessio~Cardillo\inst{1,2}\fnmsep\thanks{\email{alessio.cardillo@ct.infn.it}} \and
        Massimiliano~Zanin \inst{3,4,5}\fnmsep\thanks{\email{massimiliano.zanin@ctb.upm.es}} \and
        Jes\'us~G\'omez-Garde\~nes\inst{1,2}\fnmsep\thanks{\email{gardenes@gmail.com}} \and
        Miguel~Romance\inst{3,6}\fnmsep\thanks{\email{miguel.romance@urjc.es}} \and
        Alejandro~J.~Garc\'ia~del~Amo\inst{3,6}\fnmsep\thanks{\email{alejandro.garciadelamo@urjc.es}} \and
        Stefano~Boccaletti\inst{3}\fnmsep\thanks{\email{stefano.boccaletti@fi.isc.cnr.it}}}

\institute{Department of Condensed Matter Physics, University of Zaragoza (Spain) \and
           Institute for Biocomputation and Physics of Complex Systems (BIFI),\\ University of Zaragoza~(Spain) \and
           Center for Biomedical Technology (CTB), Technical University of Madrid (Spain)  \and
           INNAXIS Foundation and Research Institute, Madrid (Spain) \and
	Departamento de Engenharia Electrot\'ecnica, Universidade Nova de Lisboa (Portugal) \and
           Department of Applied Mathematics, Rey Juan Carlos University, Madrid (Spain)}

\abstract{
We study the dynamics of the European Air Transport Network by using a multiplex network formalism. We will consider the set of flights of each airline as an interdependent network and we analyze the resilience of the system against random flight failures in the passenger's rescheduling problem. A comparison between the single-plex approach and the corresponding multiplex one is presented illustrating that the multiplexity strongly affects the robustness of the European Air Network. 
}

\maketitle

\section{Introduction}
\label{intro}

In the last century, the application of aeronautics to the transportation of people and goods has witnessed an uninterrupted growth \cite{Heppenheimer98}. In less than a hundred years we have moved from a sparsely connected system, to a redundant one capable of moving $2.7$ billion passengers in 2011. During the last decade, scientists have studied the properties of airline transportation systems by means of network theory, unveiling their structural characteristics as done with other natural and technological complex networks. Along this period, complex networks \cite{Boccaletti06,Albert02} have extensively been used to model and understand the structures of relations beyond many real-world systems \cite{Costa11}, but only recently have some limitations of this approach been highlighted. One of the most important limitations refers to the multi-layer nature of real-world systems: nodes usually belong to different {\it layers} at the same time, and may have different neighbourhoods depending on the layer being considered. It is clear that nodes, in some complex systems, often have interactions of different kinds, which take place upon several interacting networks, i.e. constituting a so-called multiplex network.  An example of this concept is represented by social networks \cite{Wasserman}. Traditionally, social networks have been modelled as simple graphs; yet, it should be noticed that each node, representing an actor in the social network, may have different types of connections with other nodes, such as friendship, professional relationships etc. For such kind of systems, a multiplex model fits better the real situation, as it can better catch the different dynamics developing in each layer: for instance, usually the information transmitted to friends will not be the same as the one shared with colleagues. Therefore, in order to understand how the structure is affecting the global dynamics of a system, it is of utmost importance to take into account the presence of interactions at multiple levels \cite{Kurant06}. There are other concepts strongly related to the multiplex networks that have recently been introduced in the literature, such as interacting \cite{Leicht}, interdependent \cite{Buldyrev10} and multilevel networks \cite{Criado}.

Recently, several works have focused on the vulnerability of networks to cascading failures, and especially how a multi-layer structure effectively reduces the resilience of the system. For instance, Ref. \cite{Kurant07} analyzes different communication and transportation networks, composed of two layers: a physical and a logical network, the latter representing the flows of information and people. In Ref. \cite{Buldyrev10}, the Italian power grid and the Internet network are modeled as a single dual-layer system; the interconnections between both layers drammatically increase the vulnerability of the system, as a failure of a node may propagate to the other layer and generate a cascade dynamic. In Ref. \cite{Brummitt12}, a generalization of the threshold cascade model is studied, in which nodes are deactivated too if at least a given fraction of the neighbors have been deactivated. The generalization consists in introducing a multi-layer structure, that was not considered as part of the original model \cite{Watts02}: thanks to that, some topologies that were initially stable generate cascade dynamics when connected in a
multi-layer paradigm.

In this contribution, we tackle the problem of the resilience of the Air Transport Network (ATN) from a multi-layer point of view, against the deletion of a connection, that is, the cancellation of a flight. The ATN is clearly one of the tenets of our societies. In 2010, the global air transport dealt with $2.4$ billion passengers and $43$ million tonnes of cargo, has been responsible for $32$ million jobs, $2\%$ of global carbon emissions and $\$ 545$ billion in revenue \cite{IATA10}. It embraces the whole world and tightly links together the different regions, with all their individual differences.

The importance of the ATN is especially relevant when its dynamics is disturbed by external events; even when these events have only a local impact, like, for instance, a thunderstorm that forces the cancellation of a few flights, the indirect consequences (in terms of delays, passengers loosing connections, and so forth) may affect the overall performance of the system. This situation is expected to worsen in the future, as forecasted growth rates (about $5\%$ per year \cite{EUROCONTROL10}, with crises, like the WTC attack, SARS or the financial crisis \cite{Airbus09}, only having a temporary impact) will imply a tightening of the room for manoeuvre available to cope with such disturbances. The relevance of the resilience of ATN has recently been recognized in the policy-making context, as, for instance, in the European Commission's new roadmap (White Paper) to a Single European Transport Area for 2050  \cite{Eur11,Eur11b}.

The dynamics and resilience of the ATN has already been studied in the past by considering the usual single layer network formalism in which all the connections between airports are considered to be equivalent  \cite{Sridhar08,Lacasa09}. Yet, a study of ATN under the multi-layer approach is still missing. The intrinsic multi-layer nature of ATN is validated by the fact that passengers cannot use all the possible sequences of links between the airports bypassing the cost associated to the use of different airlines. This may negatively affect the resilience of the system, as well as the tools available to the system to reduce the impact of failures on the flow passengers. This demands for a study in which network and transportation sciences tackle the influence that the multi-layer architecture of the ATN has in its robustness under random failures.

\section{The European ATN as a multilayer network}
\label{network}

\begin{figure}[ht]
\begin{center}
\resizebox{0.90\columnwidth}{!}{ \includegraphics{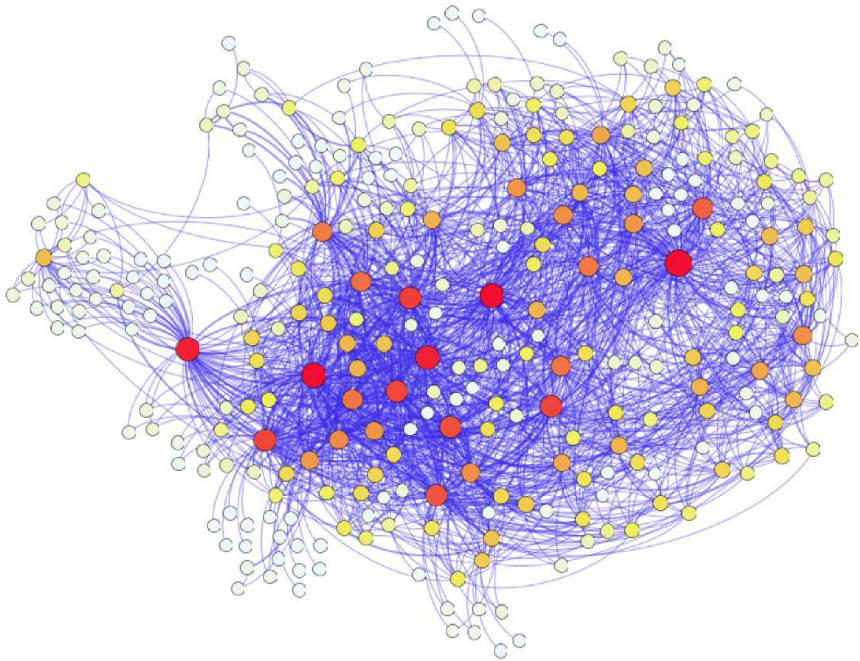} }
\caption{The European Air Transport Network (ATN). The network has been constructed by considering only commercial (both regular and charter) flights operated between two European airports the $1^{st}$ of June 2011. Size and color of nodes accounts for their degree.}
\end{center}
\label{fig:Global_network}
\end{figure}

We start by describing the structural multiplex backbone of the European Air Transport Network (ATN) considered in our model. We consider a set of 15 layers, each of which representing one of the 15 biggest airline companies operating in Europe. In each layer $\ell$ (that represents an airline $A$), the set of nodes corresponds to the set of airports operated by the airline $A$ and the links (denoted by $(i,j;\ell)$) are the flights between the airports $i$ and $j$ that are operated by airline $A$. Data corresponds to commercial IFR (Instrumental Flight Rules) operations for the $1^{st}$ of June 2011. The resulting multi-layer network (see Fig. \ref{fig:Global_network}) is an undirected system, composed of 15 layers and 308 nodes, corresponding to the $20\%$ of the operations in the European airspace.

\begin{figure}[h]

\begin{center}
\resizebox{0.3\columnwidth}{!}{ \includegraphics{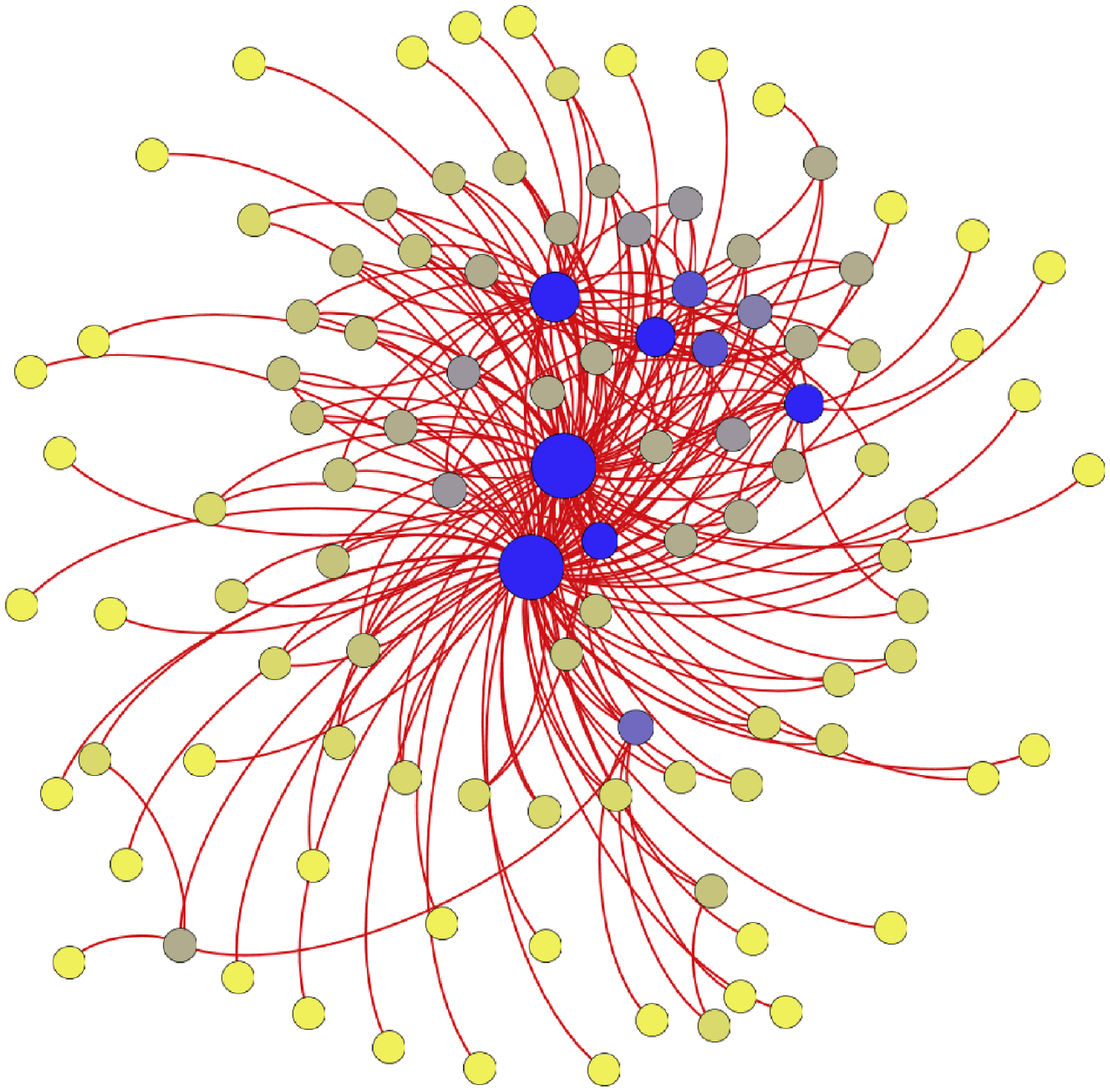} }\qquad
\resizebox{0.3\columnwidth}{!}{ \includegraphics{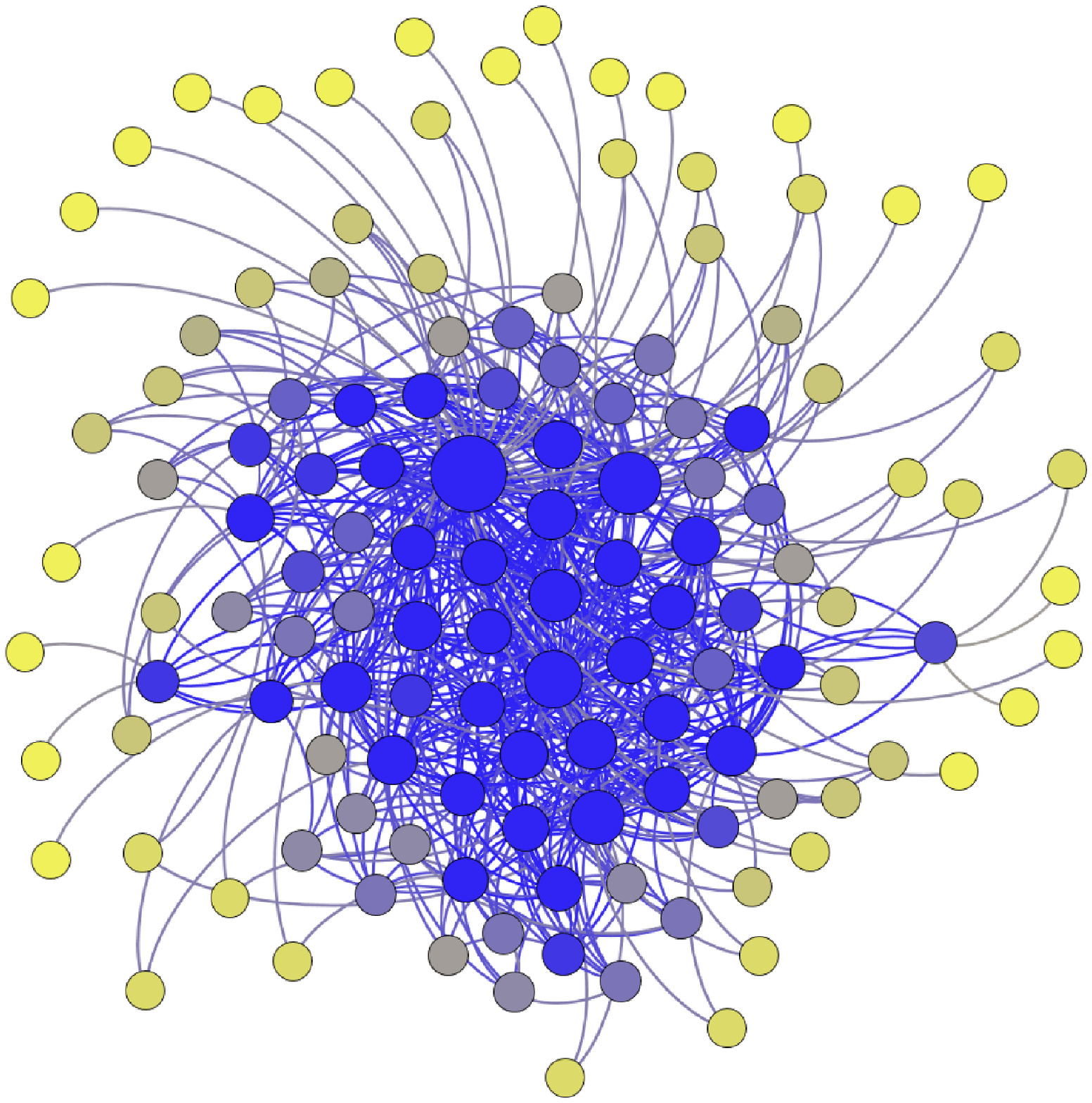} }
\caption{The Air Transport Networks of a traditional major company (on the left) and of a low-cost airline (on the right). The hubs of each network are indicated by blue big circles.}
\end{center}
\label{fig:Company_networks}
\end{figure}

Looking at the structural properties of the different layers, we realize that they are organized in two main families: (1) networks corresponding to {\it major} airlines (such as Lufthansa, Air France, or Iberia), with a scale-free networks, with hubs representing the airline headquarters; and (2) networks corresponding to the so-called {\sl low-cost} (or {\it low-fares}) airlines, with a more uniform structure due to a {\it point-to-point} organization of their business \cite{Chou90}. Figure \ref{fig:Company_networks} illustrates two of such layers: a traditional major company on the left, and of a low-cost company on the right. In this figure the hubs of each network are indicated by blue big circles; notice that the heterogeneity is much stronger in traditional companies.

\begin{figure}[b]
\begin{center}
\resizebox{0.4\columnwidth}{!}{ \includegraphics[angle=-90]{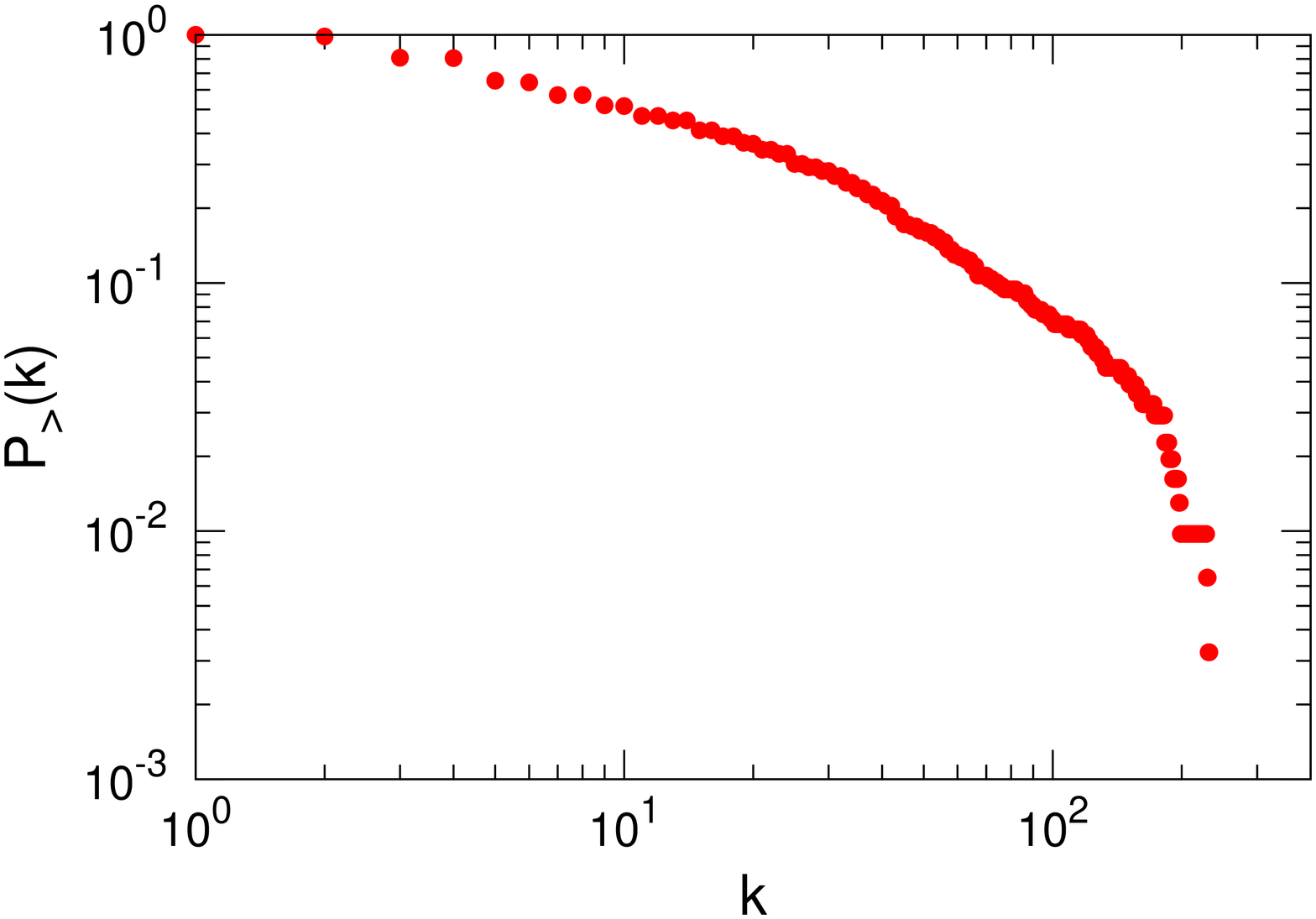} }
\resizebox{0.4\columnwidth}{!}{ \includegraphics[angle=-90]{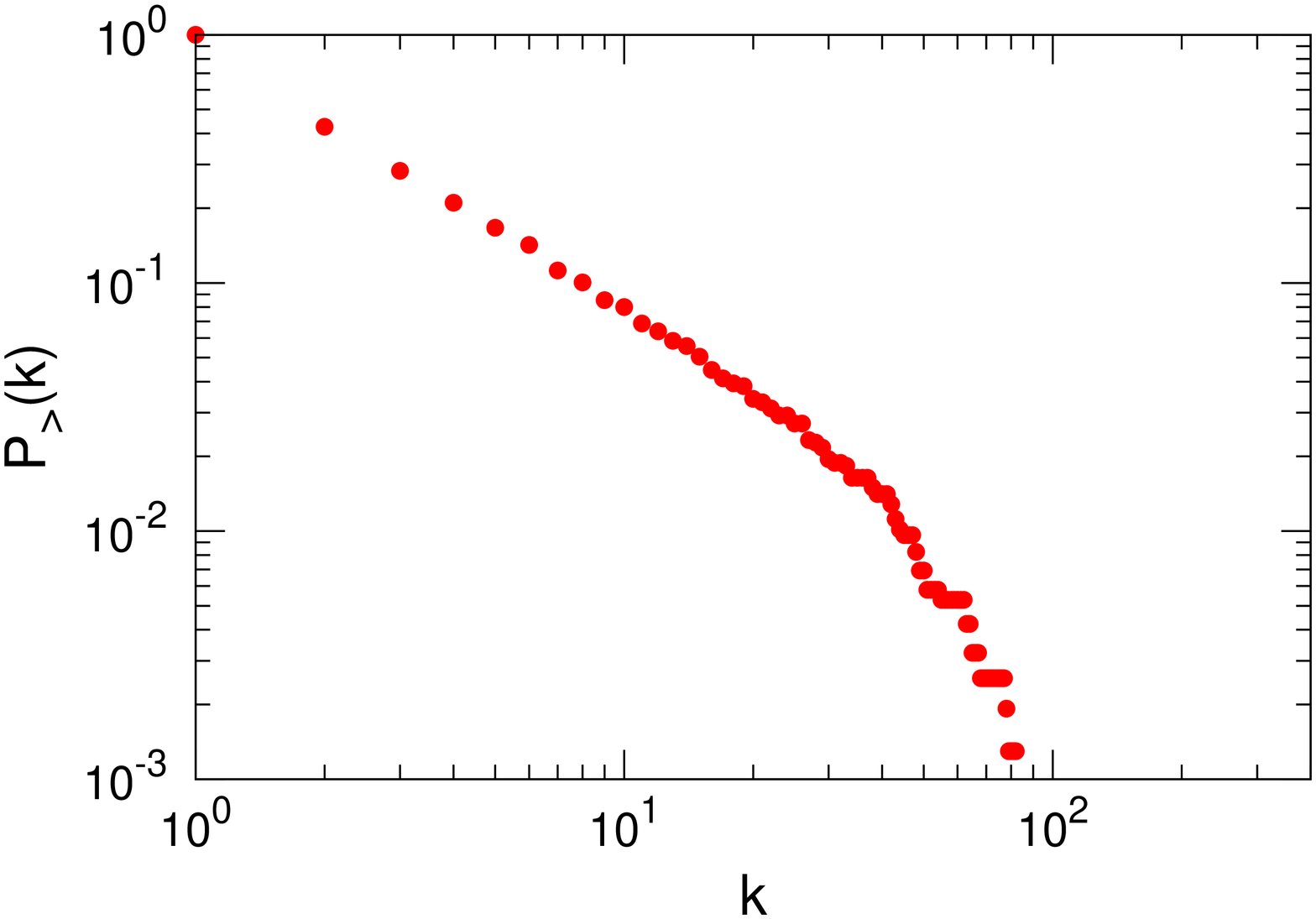} } \\
\resizebox{0.4\columnwidth}{!}{ \includegraphics[angle=-90]{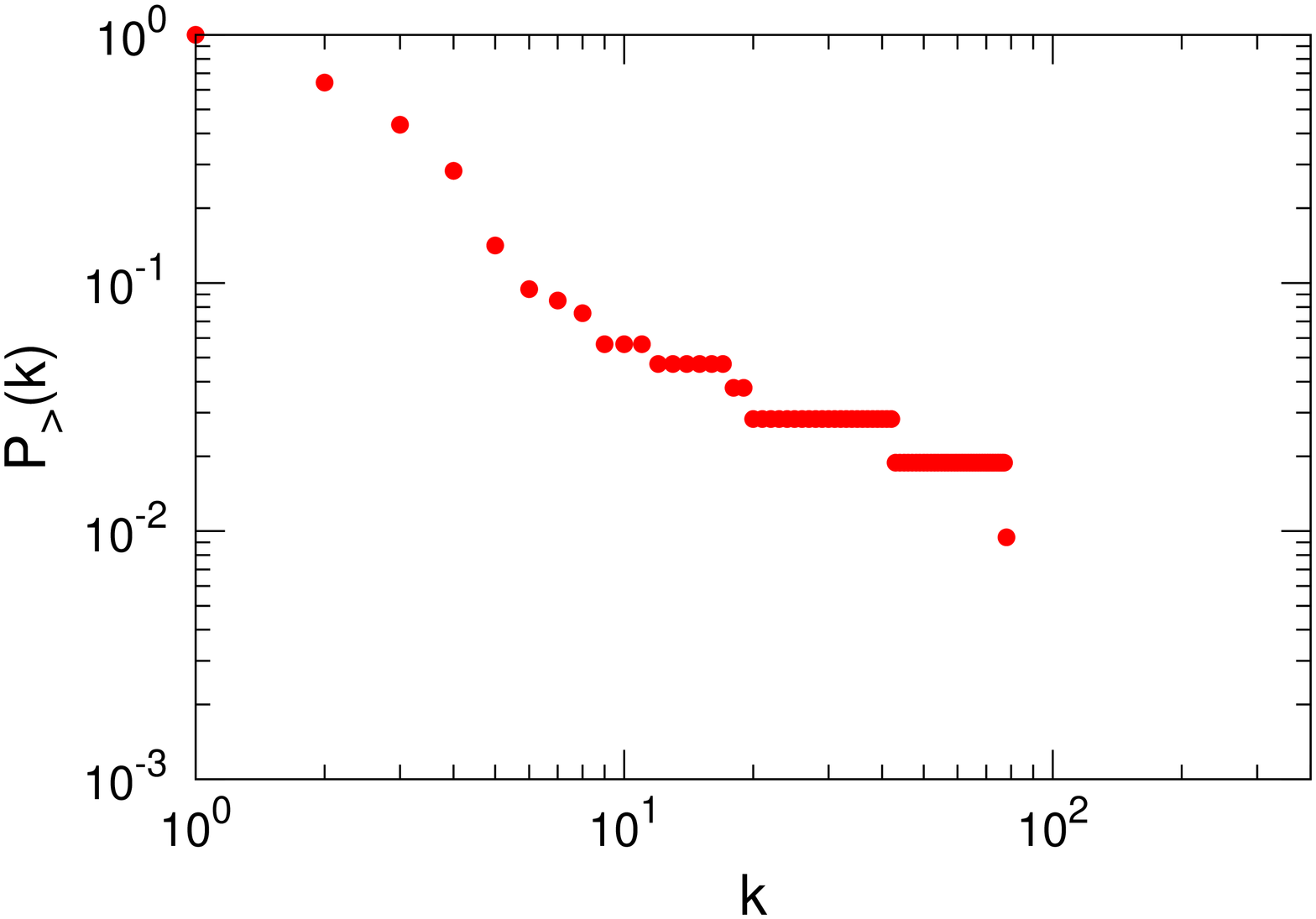} }
\resizebox{0.4\columnwidth}{!}{ \includegraphics[angle=-90]{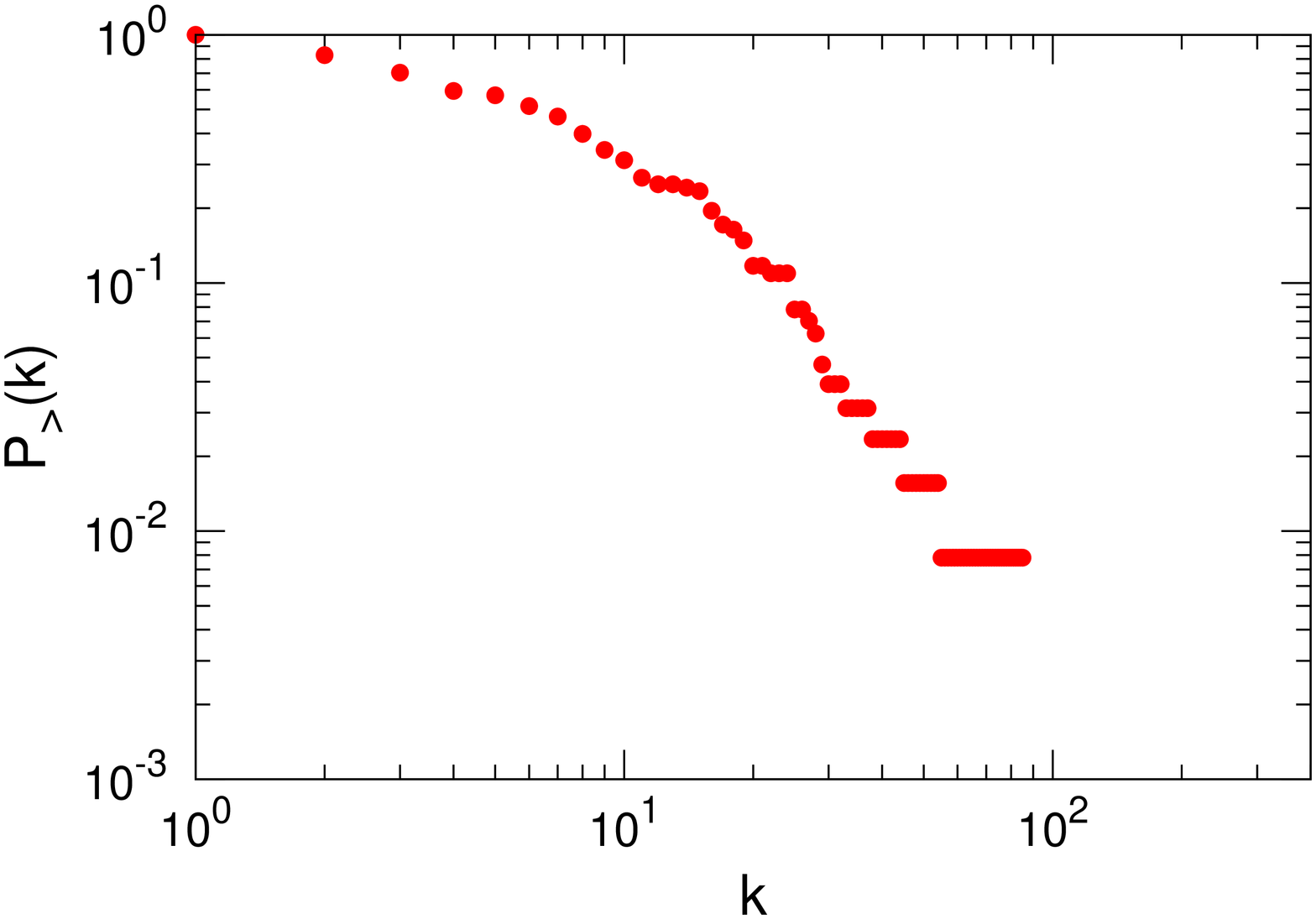} }
\caption{Example of different cumulative degree probability distributions $P_{>}(k)$ for the European ATN in $\log$-$\log$ scale. Top panels show the degree distribution for the multi-layer network model (on the left), and the average of the degree distributions of the $15$ airlines under study (on the right). Bottom panels illustrate the cumulative distributions for a single traditional major company of 106 nodes (on the left) and for a low-cost company of 128 nodes (on the right).}
\end{center}
\label{fig:acumulate}
\end{figure}

The introduction of a multiplex-type network for the European ATN produces structural properties that differ from the corresponding single-mode network, i.e., the single layer projection of the transport network. For instance, we focus here on the global degree distribution $P(k^A)$ of the multi-layer network; the global degree of each node $i$ (denoted by $k_i^A$) is calculated as the sum of the number of connections of that node over all the layers. Therefore, $k_i^A$ is defined as: 

\begin{equation}
k_i^A=\sum_{\ell=1}^Lk_i^\ell,
\label{eq:degree}
\end{equation}

where $k_i^\ell$ is the degree of node $i$ in the layer $\ell$. Fig. \ref{fig:acumulate} illustrates the cumulative probability distribution of degrees of the European ATN in $\log$-$\log$ scale. Clearly, there are strong differences between the distribution for the multi-layer network model (top left panel) and the average of the cumulative degree distribution over all the layers considered (top right panel). Note that the degree of nodes in the case of the multi-layer model is greater than the corresponding degree in the classic one-layer approach. This phenomenon is even more explicit in the case of the hubs, since a link than could happen in different layers is counted as many times as it is present in the multi-layer network, while it is only counted only once in the classic model. Despite this fact, one could expect that this enhancement of the degree is uniform along the network, but the real situation is quite far from this. The heterogeneity of the structure and distribution of each layer makes that the effects of the enhancement of the degree of each node in the multi-layer network is very disperse and therefore the degree distribution in the multi-layer model is very different from the corresponding classic model. Furthermore, if we compute the average degree distribution along all the layers in the network (see the top right panel in figure~\ref{fig:acumulate}), the result is quite different from the corresponding figure for the multi-layer model. Note that the average degree distribution illustrate the average degree distribution if we pick up a layer at random and we look at its degree distribution. Hence, the significant differences between the top panels in figure~\ref{fig:acumulate} illustrate the different behavior of the multiplex model and the corresponding for each single layer, that comes from the heterogeneity of the structure and distribution of the network. A similar situation occurs if we consider the cumulative probability distribution of each single layer individually (see Fig. \ref{fig:acumulate} Bottom). 

\section{The model}
\label{model}

As anticipated above, we will consider the set of direct flights of the same airline as the links of one independent network, {\em i.e.}, a single layer. On the other hand, each of the $N$ nodes of the ATN will be present in each of the layers. Thus the collection of the $L$ layers composes a multiplex representation of the ATN. Each of the layers will be denoted by a super index $\ell=1,\ldots,L$ so that the shortest distance between each couple of nodes within the same layer is denoted as $d_{ij}^\ell$ and the degree of a node $i$ within layer $\ell$ is $k_i^\ell$.

Once the topology of each layer of the multiplex ATN is characterized, we implement a model for the flow of a set of $N_p$ passengers. First we assign the routes followed by each of the $N_p$ passengers that move across the ATN. To this aim, and for each of the $N_p$ passengers, we randomly choose two nodes of the ATN (one accounting for the origin and one for the destination of the passenger). Both nodes are selected proportionally to their global degrees $k_i^A$, as defined in Eq. \ref{eq:degree}. In this way, a node $i$ will be selected as origin or destination of a given passenger with a probability:
\begin{equation}
P(i)=\frac{ k_i^{A}}{\sum_{j=1}^Nk_j^{A}}\,.
\end{equation}
Obviously, paths starting from and ending at the same node are not allowed. Once the origin and destination of a passenger have been chosen, we search among all the layers the one for which the distance between the origin $i$ and the destination $j$ is minimal, {\it i.e.} $d_{ij}=\min\{d_{ij}^\ell, \ell=1,\ldots,L\}$. 
The distance between two airports, $d_{ij}$, is here defined as the hopping distance, i.e. the sum of the number of {\it jumps} needed to reach the destinations; other factors usually taken into account by the passenger (like duration of the flight, cost, and so forth) are here disregarded.
If there is more than one layer with the same minimum distance, one of them is randomly selected with equal probability;
notice that this is equivalent to a passenger selecting one of the multiple airlines available to reach its destination. Finally, after the layer is selected, we compute the shortest path between origin and destination; a shortest path is randomly chosen when more than one was available.

The above process ends when all the $N_p$ passengers have selected a couple of nodes (that is, their origin and destination), an airline (the layer) and a route (the shortest path between origin and destination nodes in the selected layer). Then, we can compute the load of each link $(i,j;\ell)$, $L(i,j;\ell)$, in each of the layers of the ATN, defined as the number of passengers whose path pass through it. In addition, for each link in the system, we assign a \emph{maximum load} $L^M(i,j;\ell)$ as:
\begin{equation}
L^M (i,j;\ell)= L(i,j;\ell)(1+ f_{tol})\;,
\end{equation}
where $f_{tol}$ accounts for the fraction of additional load that each link can handle. For instance, a value of $f_{tol}=0.2$ implies that airlines leave a number of vacant seats equal to the $20\%$ of the real load. In what follows, we analyze situations in which $0 \leq f_{tol} \leq 0.3$, in line with the load factors observed in real operations ($70\%$ for short flights, and $80\%$ for long-range connections \cite{AEA}).

Once the model has been initialized, we simulate a random failure of the system by randomly removing a fraction of the links. With this aim, we visit each link connecting two nodes $i$ and $j$ in a given layer $\ell$, and with some probability $p$ we remove that link. As a consequence, all the passengers whose original paths passed through one (or more) of the removed links have to be re-scheduled, {\em i.e.}, they are forced to look for an alternative route between their departure and destination airports. As a previous step to the re-scheduling of a passenger, we decrease by one the load of the remaining active links in the passenger's original path.

\subsection{Re-scheduling algorithm}

After simulating the perturbation of the original system, we proceed with the re-scheduling phase. For each affected passenger, we try to find a new path between the origin $i$ and the destination $j$ of distance $d_{ij}(n)=d_{ij}+n$ (being $d_{ij}$ the original distance in the unperturbed ATN), with $n=\{0,1,2,\ldots\}$. Obviously, we start by trying to allocate passengers in new routes with $n=0$, so that the number of connections required for completing the trip (a proxy for the cost incurred by passengers) is not increased. Thus, for a given value of $n$, we proceed as follows:

\begin{itemize}
\item[{\em (i)}] We recalculate all the {\em active paths} between each pair of nodes $(i, j)$, within the same layer $\ell$. We impose the distance between the pairs of nodes to be $d_{ij}(n)=d_{ij}+n$. Two situations may lead to the absence of {\em active paths} of length $d_{ij}$ in a layer $\ell$:
\begin{itemize}
\item[(a)] there are no paths of this length in the original multiplex graph.
\item[(b)] there are some paths of length $d_{ij}(n)$, but all of them contains removed or full (see below) links.
\end{itemize}
\smallskip
After this stage, each passenger is classified as either {\em fly} (he/she already has a route assigned, not affected by the removal of links),  {\em re-scheduling} (he/she has the possibility of being assigned to an active route)  or {\em no-fly} (there is no active path of distance less or equal to $d_{ij}(n)$ in any of the layers).
\item[{\em (ii)}] We take all the passengers one at the time in the {\em re-scheduling} group. For each of them:
\begin{itemize}
\item[(a)] We take their original layer and try to construct an (active) alternative path enabling the passenger to reach its destination whenever possible. If the chosen active path does not contain any full (see below) link, then the passenger is classified as {\em fly} and the load of each link in the chosen path is increased by one. Those links that reach their total capacity $L^M(i,j;\ell)$ with the addition of this last passenger are then classified as full.
\item[(b)] If after step ({\em ii}.a) the passenger remains as re-scheduling, we repeat this last step for all the layers that contain at least one active path of length $d_{ij}(n)$ between its origin $i$ and destination $j$. Again, if the passenger is successfully re-scheduled, it goes to the {\em fly} club and we add $1$ to the load of each links used. If any of these links reaches its capacity $L^M(i,j;\ell)$, then it is classified as full.
\end{itemize}
\item[{\em (iii)}] Once all the passengers in the {\em re-scheduling} compartment have been processed, we check the remaining number of {\em re-scheduling} passengers. If it is non-zero,  we go again to step {\em (i)}.
\end{itemize}

At the end of the above iterative process, we partitioned the set of passengers into the subsets of {\em fly} and {\em no-fly}. We then perform the above process for different values of $n$ increasing from $n=0$. After each round $n$, passengers classified as {\em no-fly} are introduced again in the model as {\em re-scheduling} at the beginning of round $(n+1)$. In principle, $n$ can be increased as many times as desired; nevertheless, to be realistic, we stop the algorithm at $n=2$, meaning that passengers could, at most, look for alternative paths that are up to two steps longer than their original ones. Thus, at the end of round $n=2$, {\em no-fly} passengers are those for which no active path of the former length exists between their origins and destinations. The rest of the passengers have been efficiently re-scheduled and take part of the final {\em fly} club\footnote{Let us remark that we are assuming that passengers try to move to other airlines (layers) in order to avoid longer trips than those originally planned, {\em i.e.}, the case $n=0$. Only when this latter attempt fails, they consider to perform longer trips ($n>0$).}.
It is worth noticing that this re-scheduling algorithm does not include any information about alliances between airlines; this means that (i) passengers cannot plan their trip by connecting flights of different airlines, and (ii) that the re-scheduling is unbiased, while in the real world airlines try first to accomodate passengers in flights of the same alliance.

\section{Results}
\label{results}

In this section we will explore the effects that link deletion causes on the flow of passengers across the multiplex ATN. To shed light on the effects on multiplexity, we compare the results obtained in the multiplex network with those of the aggregate ATN. The model introduced in the previous section has two parameters, namely the probability $p$ that a link is deleted, and the fraction of tolerance $f_{tol}$ that airlines assign to their connections. In what follows, we explore the robustness of the ATN as a function of the former two parameters.

\subsection{Robustness of the ATN as a multiplex network}

In order to characterize the effects that link deletion has on the re-organization of the flow in the multiplex ATN, we considered the partition into different groups of the total population of $N_{A}$($\leq N_p$) passengers affected by link deletion. This population can be divided into two groups: one, of size $N_{f}$, composed of those passengers that can reach their destination thanks to the re-scheduling process; and another group, of size $N_{nf}$, composed of those passengers that cannot be accommodated after the random failure of the system. Following this classification, we have that $N_{A}=N_{nf}+N_{f}$. In order to clearly monitor the effect of having different layers (airlines) in the multiplex ATN, we further split the group of $N_{f}$ passengers that have been successfully re-scheduled into two other groups: those $N_{sl}$ passengers that are re-scheduled within the same layer as originally planned, and those $N_{ol}$ passengers that were forced to change layer in order to reach their corresponding destinations. With this new division we obtain the following equality: $N_{A}=N_{nf}+N_{sl}+N_{ol}$.

\begin{figure}
\begin{center}
\resizebox{0.98\columnwidth}{!}{ \includegraphics{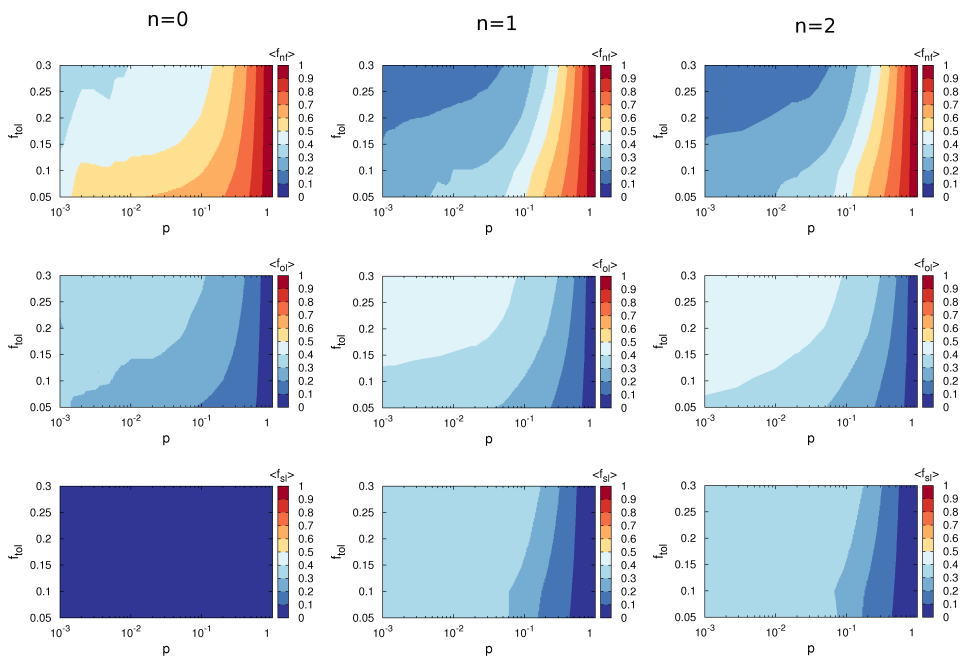} }
\caption{Outcome of the re-scheduling process as a function of the probability of link failure, $p$, and load tolerance, $f_\text{tol}$. Each column displays (from top to bottom): the average fraction of passengers that cannot fly ($f_{nf}$), that of those that are re-scheduled in other layers ($f_{ol}$), and that for those re-scheduled within the original layer ($f_{sl}$). Each column accounts for the possibility of scheduling passengers on paths with length up to $d_{ij}(n)$ with $n=0$ (left), $n=1$ (center), and $n=2$ (right). Results shown here refer to a population of  $N_{p} = 50000$ passengers and are averaged over 50 different realizations. $p$ spans logarithmically in the range $[10^{-3},1]$, while $f_\text{tol}$ spans in the range $[0,0.3]$.}
\end{center}
\label{fig:re-scheduling-50000-d2}
\end{figure}

The three groups ({\em no-fly}, {\em same-layer} and {\em other-layer}) completely describe the final state of the population of affected passengers. In Fig.~\ref{fig:re-scheduling-50000-d2}  we plot the fraction of passengers belonging to each compartment: $f_{nf}=N_{nf}/N_{A}$, $f_{ol}=N_{ol}/N_{A}$ and $f_{sl}=N_{sl}/N_{A}$, as a function of the two parameters $p$ and $f_{tol}$. We also show how these quantities behave by iterating the re-scheduling algorithm for several values of $n$. Namely, in the left column of the figure we show (from top to bottom) the panels corresponding to $f_{nf}(p,f_{tol})$, $f_{ol}(p,f_{tol})$ and $f_{sl}(p,f_{tol})$ for $n=0$, {\em i.e.}, when passengers are allowed to perform alternative trips only if their lengths are equal to the original one. In this plot we observe that there are almost no re-scheduled passengers flying across their original layers. This points out the low degree of redundant shortest-paths between two nodes in a given layer. As a consequence, almost all the successfully re-scheduled passengers are forced to change airline. The number of efficiently re-scheduled passengers ($N_{sl}+N_{ol}$)  decreases with $p$ and increase with $f_{tol}$. However, as can be observed in the top panel, the number of {\em no-fly}-passengers is extremely large even for a low rate of link deletion and a high degree of tolerance. Namely, for a value of $p\sim 10^{-2}$ and a degree of tolerance of about $10\%$, the fraction of {\em no-fly} passengers lies over the $50\%$ of the population initially affected by the removal of links.

The constraint imposed in the case $n=0$ seems too restrictive to achieve an efficient re-allocation of passengers, as it does not allow passengers to perform alternative paths in their respective original layers, at the cost of increasing the total length. Therefore, we relax this constraint and explore the cases $n=1$ and $n=2$ in the middle and right columns respectively.
From these panels we observe that the average fraction of {\em no-fly} passengers (upper panels) is much lower than in the previous case $n=0$. The decrease becomes more apparent for those regions corresponding to high values of load tolerance and low values of $p$. Remarkably, contrary to the case for $n=0$, both for $n=1$ and $n=2$ some of the re-scheduled passengers succeded in traveling through alternative routes within their original layer. Besides, we observe that the plots corresponding to $n=1$ and $n=2$ are quite similar, pointing out that allowing the search for routes with $n>2$ would not improve the results. Therefore, the plot of $f_{nf}(p,f_{tol})$ for $n=2$ indicates a relative weakness of multiplex ATN with respect to perturbations, given that, even for very large values of tolerance and very low values of $p$, there is always some non-zero fraction of {\em no-fly} passengers.

\subsection{Aggregate network results}

In order to gain more insight on the effects of the multiplex structure of our system, we now show the results obtained with the same re-scheduling algorithm on the aggregate version of the ATN. Such aggregate network is obtained by merging all the layers of the multiplex representation into a single one, {\em i.e.} by projecting the multiplex graph into a simplex one. This projection produces a complex network with the presence of multiple links between those couples of nodes that were connected in more than two layers; in other words, the number of connections between two airports is given by the number of airlines operating between them. In order to test whether the robustness of the aggregate network is larger than that of the multiplex network, we have performed the same link removal process followed by the re-scheduling program described in Section \ref{model}, this time considering the single layer comprising all the links in the aggregate ATN.

\begin{figure}
\begin{center}
\resizebox{0.98\columnwidth}{!}{ \includegraphics{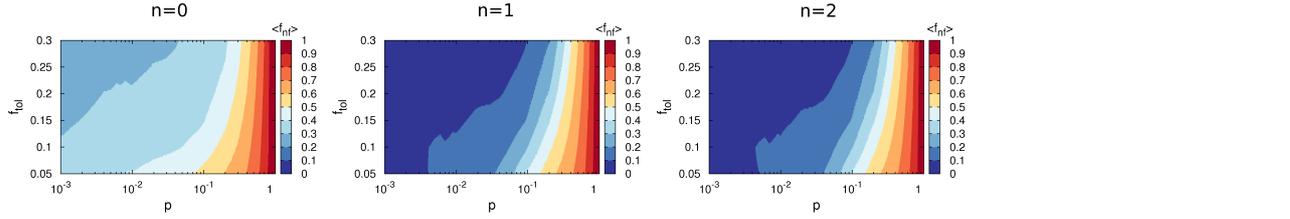} }
\caption{Effect of link removal on the final state for the case of the aggregate network. We plot the average fraction of passengers that are not able to fly ($f_{nf}$) as a function of the probability of link removal, $p$, and the load tolerance, $f_{tol}$. Each column accounts of the possibility of re-scheduling passengers by means of paths of length up to $d_{ij}(n)$ with $n=0,1,2$. Simulations are run with the same parameters and under the same conditions of those shown in Fig.~\ref{fig:re-scheduling-50000-d2}.}
\end{center}
\label{fig:re-scheduling-aggregate-50000-d2}
\end{figure}

Fig.~\ref{fig:re-scheduling-aggregate-50000-d2} shows the final state of the system for the same three scenarios explored for the multiplex ATN, namely $n=0$, $1$ and $2$. Since the aggregate ATN is composed of a single layer, in this case we only focus on the fraction of passengers affected by link deletion that are not able to be efficiently re-scheduled, $f_{nf}(p,f_{tol})$. As observed from the three panels in Fig.~\ref{fig:re-scheduling-aggregate-50000-d2}, compared with the corresponding panels $f_{nf}(p,f_{tol})$ for the multiplex ATN, the fraction of {no-fly} passengers decreases considerably in the three studied cases. In particular, while for those regions of the plot corresponding to $p>10^{-1}$ remains roughly the same as in the case of the multiplex ATN, the main differences show up for low values of p; specifically, for $n>0$ we can observe regions for which almost all the affected passengers can be re-scheduled, with an almost empty {\em no-fly} set. Again the panels corresponding to $n=1$ and $n=2$ are identical pointing out that the system is unable to achieve a better balance of affected passenger by increasing the length of the alternative trips. As a conclusion, the aggregate network shows an improved robustness with respect to the multiplex one, and a null impact of link deletion for some range of parameters.

This comparison confirms that multiplexity affects the robustness of the ATN. The root of the differences between the performance of both topologies is the constraint imposed by the multiplex architecture, which forces passengers to move within single layers. Therefore, in order to find an efficient alternative path, the affected passenger cannot mix connections of different airlines (layers) into the same path, thus reducing his capacity of optimizing the movement. This constraint disappears in the aggregate network, allowing affected passengers to make use of {\em hybrid} alternative paths. This provides the aggregate system way out to re-schedule the affected population of passengers in an efficient manner.

\section{Conclusions}
\label{conclusions}

We presented a model for studying the re-scheduling problem in the European Air Transport Network using the paradigm of multiple layers structure, where each layer is made by the flights of a given airline. it is worth noting that, when comparing this multi-layer network with an equivalent single layer representation, topological characteristics differ, both qualitatively and quantitatively, as exposed in Section \ref{network}. Furthermore, on top of this multiplex network, we built a dynamical model, accounting for the re-scheduling problem of a group of passengers affected by the random failures of a set of connections. The affected passengers are then re-scheduled on new itineraries according to the availability of new routes (and free seats) in their former airline first, or eventually in a different one. The availability of routes is modulated by the probability of link failure $p$, and the tolerance on the load of a link $f_{tol}$. We presented our results in terms of the number of those passengers that are successfully re-scheduled and those for which the re-scheduling procedure fails. To achieve a deeper insight on the effects of dealing with a layered structure, we further subdivided passengers who are successfully re-scheduled into two subcategories: those which continue their trip using the same airline and those who, instead, are forced to switch to a different one. In addition, in order to increase the realism of our model, we allowed passengers to be re-allocated also on paths which are longer than the former ones. When compared to those corresponding to the single-layer representation, our results indicate that the multi-layer structure strongly reduces the resilience of the system against perturbations. In other words, the use of a {\it projection} of the ATN system is an over-simplification that results in an over-estimation of the resilience of the ATN. While it is known that a multi-layer structure can drammatically change the resilience of the system \cite{Buldyrev10}, to the best of our knowledge this is the first application of such representation to the air transport system; all previous studies (see, for instance, \cite{Lacasa09}) only considers projections of the network. We anticipate that this framework may be an important tool for policy makers in the near future, especially when other elements (e.g., more real estimation of the distance between airports, airline alliances, estimation of the costs of re-routing, etc., here excluded for the sake of simplicity) would be included.
We also believe that these results could also be valid in other real-world complex systems, which have been widely studied in the last decade under the single layer network paradigm, when their multiplex nature is taken into account.

\section*{Acknowledgements}
This work has been partially supported by the Spanish DGICYT
under projects FIS2008-01240, MTM2009-13848 and FIS2011-25167; by the Comunidad
de Arag\'on through Project No. FMI22/10. J.G.G is supported by MICINN through the
Ram\'on y Cajal
program.
Authors gratefully acknowledge EUROCONTROL and the ComplexWorld Network (\url{www.complexworld.eu}) in the context of the SESAR Work Package E for sharing the operational data set.
The paper reflects only the authors' views. EUROCONTROL and the ComplexWorld Network are not liable for any use that may be made of the information contained therein.
Authors also thank David Papo for his suggestions and corrections on the Manuscript.

\end{document}